\documentclass[10pt,twoside,onecolumn]{article}
\usepackage[]{latexsym,amsmath,amssymb}
\usepackage[]{epsfig}
\newcommand{\be}{\begin{eqnarray}}
\newcommand{\ee}{\end{eqnarray}}
\newcommand{\ud}{\mathrm{d}}

\newcommand{\lp}{\ell_{\rm p}}
\newcommand{\mpl}{m_{\rm p}}

\newcommand{\expec}[1]{\mbox{$\langle #1\rangle$}}

\title{\bf Gravitational renormalization of quantum field theory}
\author{Roberto Casadio\thanks{casadio@bo.infn.it}
\\
Dipartimento di Fisica, Universit\`a di Bologna,
and I.N.F.N.,
\\
Sezione di Bologna, via~Irnerio~46, 40126~Bologna, Italy}
\begin{document}
\maketitle
\begin{abstract}
We propose to include gravity in quantum field theory non-perturbatively,
by modifying the propagators so that each virtual particle in a Feynman graph
move in the space-time determined by the four-momenta of the other particles in the same graph.
By making additional working assumptions, we are able to put this idea at work in a simplified
context, and obtain a modified Feynman propagator for the massless neutral scalar field.
Our expression shows a suppression at high momentum, strong enough to entail finite
results, to all loop orders, for processes involving at least two virtual particles.
\end{abstract}
%
%
%
%
\section{Introduction}
\setcounter{equation}{0}
Pauli, long ago~\footnote{See, e.g., in Refs.~\cite{pauli-a,pauli-b}.},
suggested that gravity could act as a regulator for the ultraviolet (UV) divergences
that plague quantum field theory (QFT) by providing a natural cut-off at the
Planck scale.
Later on, classical divergences in the self-mass of point-like particles were
indeed shown to be cured by gravity~\cite{adm}, and the general idea has since then
resurfaced in the literature many times
(see, {\em e.g.}, Refs.~\cite{deser,dewitt,salam,ford,woodard,woodADM}).
In spite of that, Pauli's ambition has never been fulfilled.
\par
As it happens, QFT has been successfully used to describe particle physics in
flat~\cite{peskin} (or curved, but still fixed~\cite{birrell}) space-time,
where standard renormalization techniques work very well to produce testable results,
notwithstanding the presence of ubiquitous singularities stemming from the very
foundations of the theory, that is the causal structure of (free) propagators which
allow for too large a number of high energy modes.
We have thus grown accustomed to the idea that the parameters in the 
Lagrangian of the Standard Model (or generalisations thereof)
have no direct physical meaning and infinite contributions may be
subtracted to make sense of mathematically diverging integrals.
The modern approach to renormalization~\cite{wilson} views the occurrence
of such infinities as a measure of our theoretical ignorance of nature.
Every Lagrangian should, in turn, be considered an effective (low energy)
description doomed to fail at some UV~energy scale $\Lambda$~\cite{weinberg97},
above which the correct counting of degrees of freedom will be given
by an unknown theory.
On the other hand, gravitational corrections to the Standard Model amplitudes,
to a given order in the inverse of the Planck mass $\mpl$, are negligibly small
at experimentally accessible energies~\cite{donoghue}.
These facts briefly elucidate the main theoretical reason that makes it so
difficult to use gravity as a regulator:
if it is to provide a natural solution to the problem of UV~divergences,
gravity must be treated non-perturbatively~\cite{woodard,woodADM}.
\par
Taking a step back to the basics, we should notice that, in the QFT community,
gravity is mainly viewed as a spin-2~field, which also happens to describe distances
and angles (to some extent).
As such, the most advanced strategy to deal with it is the background
field method for functional integrals~\cite{bfm,peskin}, according to which
one expands the Einstein-Hilbert Lagrangian (or a generalisation thereof)
around all the fields' classical values, including the classical background metric.
The latter is reserved the role of defining the causal structure of space-time,
whereas the quantum mechanical part yields the graviton propagator and
matter couplings (of order $\mpl^{-2}$).
The effect of gravity on matter fields can then be analysed perturbatively by
computing the relevant Feynman graphs~\cite{veltman}.
A notorious consequence of this approach is that, by simple power counting,
pure gravity (all matter fields being switched off) is seen to be non-renormalizable,
a ``text-book'' statement by now~\cite{shomer}, which is however debated occasionally.
For example, Ref.~\cite{woodard} suggested that perturbative
expansions are performed in the wrong variables and that Einstein gravity
would  appear manifestly renormalizable if one were able to resum
logarithmic-like series~\cite{salam}.
In the physically more interesting case with matter,
non-perturbative results can be obtained in just a very few
cases, one of particular interest being the correction to the self-mass of a
scalar particle, which becomes finite once all ladder-like graphs containing
gravitons are added~\cite{dewitt}.
A remarkable approach was developed in Refs.~\cite{reuter98-a,reuter98-b},
in which a tree-level effective
action for gravity at the energy scale $\mu$ is derived within the background
field method, but without specifying the background metric {\em a priori\/}.
The latter is instead {\em a posteriori\/} and self-consistently equated
to the quantum expectation value determined by the effective action
at that scale.
This method does not involve cumbersome loop contributions and hints that
gravity might be {\em non-perturbatively\/} renormalizable~\cite{reuter08},
with the gravitational coupling showing a non-Gaussian UV~fixed point,
thus realising the {\em asymptotic safety\/}
conjectured several decades ago by Weinberg~\cite{weinberg}.
\par
More recently, an even bolder statement was put forward in Refs.~\cite{dvali-a,dvali-b},
where it was observed that all kinds of matter should ``classicalise'' at sufficiently
high energy, by simply generating black holes.
The idea is essentially based on Thorne's hoop conjecture~\cite{hoop}
(see also Ref.~\cite{hoop2} for a simple derivation):
two colliding particles of total centre-mass energy $E$ will form a black hole
whenever the impact factor $b$ is shorter than the horizon radius of the
corresponding black hole of mass $E$,
\be
b\lesssim G_{\rm N}\,E\simeq \frac{\lp\,E}{\mpl}
\ .
\ee
Since the length scale $b$ probed by the process is itself bounded by
the energy scale according to the uncertainty principle,
\be
b\sim\Delta b \gtrsim \Delta E^{-1}\sim E^{-1}
\ ,
\ee
one concludes that there should be an energy scale $\Lambda_{\rm cl}\sim\mpl$
above which the usual perturbative QFT approach simply breaks down,
because the system involves (bound) classical configurations in the final state:
increasing the energy above $\Lambda_{\rm cl}$ just makes the black hole larger
and quantum fluctuations around it weaker.
This is similar to the effective approach we outlined previously, but differs in that
it does not involve (unspecified) new physics above $\mpl$.
Of course, the central issue now becomes the description of the gravitational
collapse (of the two quantum particles into a classical black hole),
whose details are not yet fully understood at the purely classical
level~\cite{payne-a,payne-b,payne-c,payne-d},
not to mention in a QFT context~\cite{fischler-a,fischler-b,fischler-c}
(for some more recent analysis, see Refs.~\cite{hoop2,BHform-a,BHform-b}).
A better comprehension of this transition would also yield a bridge with
the asymptotic safety scenario, provided the running of the gravitational
coupling constant is fast enough to avoid the space-time singularity
predicted by GR~\cite{chsu}. 
In the lack of explicit results, there is room for speculations.
For example, at the classicalization scale $\Lambda_{\rm cl}$ one can conceive
the emergence of more exotic objects, like the so called quantum black holes~\cite{xavier}.
It also raises the question whether the quantum nature of gravity plays any role at all
in physical processes, since quantum graviton exchanges are suppressed by inverse
powers of $\mpl$ at low energy, and by black hole formation at high energy.
\par
Based on the idea that QFT is an effective approach~\cite{weinberg97},
different attempts have taken a shortcut and addressed the effects of gravity on
the propagation of matter field modes directly, {\em e.g.}, by employing modified dispersion
relations or uncertainty principles at very high (usually referred to as
{\em trans-Planckian\/}) energy~\cite{unruh-a,unruh-b}.
Some works have postulated such modifications, whereas others have tried to derive them
from (effective) descriptions of quantum gravity (see, {\em e.g.\/},
Refs.~\cite{maggiore-a,maggiore-b,maggiore-c,maggiore-d,maggiore-e}).
It is in fact common wisdom that, for energies of the order of $\mpl$ or larger, the
machinery of QFT fails and one will need a more fundamental quantum theory of gravity,
such as String Theory~\cite{string} or Loop Quantum Gravity~\cite{loop}. 
Quite interestingly, both hint at space-time non-commutativity~\cite{szabo} as an
effective implementation of gravity, with the scale of non-commutativity
of the order of the Planck length $\lp$ acting as a natural regulator
(or $\mpl$ as a covariant upper bound for the spectral decomposition,
see~\cite{kempf} and References therein).
A new feature which, in turn, follows from space-time non-commutativity is the
IR/UV~mixing, whereby physics in the infrared (IR) is affected by
UV~quantities~\cite{IRUV-a,IRUV-b}.
This feature gives us hope of probing (indirectly) such an extreme energy
realm in future experiments~\cite{das} or even using available data of very large scale
(cosmological) structures.
\par
In the present work, we shall propose a yet different strategy to incorporate gravity,
along the above line of reasoning, in a very conservative and somewhat ``minimal'' way.
Inspired by the simple semiclassical perspective in which gravity is described by Einstein's
geometrical theory and matter by perturbative QFT, gravity is not viewed as a spin-2 field
(although with very complicated interactions with itself and with matter fields),
but rather as the causal structure of space-time (or the manifestation thereof), a property the background field
method reserves to the classical part of the metric only.
The modified propagators for matter fields should therefore take into
consideration the presence of each and every source, classical or virtual, in a
given process mathematically described by Feynman's diagrams.
In this respect, our approach falls close to the asymptotic safety scenario~\cite{weinberg}
(since it does not require new physics), but short of the classicalization scheme~\cite{dvali-a}
(since it does not make any specific ansatz about the crucial role played by black holes).
Of course, philosophical perspectives aside, the relevant question is whether
this idea leads to different (or the same) phenomenological predictions with respect to
the other approaches to UV physics currently available, but we are in a fairly premature
stage to assess that.
\par
The rest of this paper is conceptually divided into two parts:
the general proposal is described in the next Section, where we briefly review the idea
of semiclassical gravity, the interplay between propagators, the causal structure
of space-time and UV~divergences,
and then list four prescriptions which should serve as guidelines in order to modify the
propagators accordingly;
in Section~\ref{scalar}, we shall try to apply the proposal and obtain a modified scalar field
propagator to estimate the UV behaviour of the four-point function to one-loop.
Let us remark that the second part of the paper is based on several more working
assumptions, in addition to the general guidelines, and the results about transition
amplitudes are therefore a consequence of both the general idea and some
simplifications which might indeed appear more questionable.
\par
We shall use units with $c=\hbar=1$ and the Newton constant $G=\lp/\mpl$.
\section{Geometrical gravity in QFT}
\setcounter{equation}{0}
\label{semi}
In order to make contact with the physics from the very start, let us note that
one should consider two basic energy scales, one related to phenomenology
and one of theoretical origin, namely:
\begin{description}
\item[a)]
the highest energy presently available in experiments, say
$E_{\rm exp}\sim 1\,$TeV, and
\item[b)]
the Planck energy $\mpl\sim 10^{16}\,$TeV.
\end{description}
It is well assessed that, for energies up to $E_{\rm exp}$, the Standard Model (without gravity)
and renormalization techniques yield results in very good agreement with the data.
Further, finite, albeit experimentally negligible, quantum gravitational corrections
can be obtained by employing the effective QFT approach~\cite{donoghue}
(which also yields some -- but not all -- of the general relativistic
corrections to the Newtonian potential).
At the opposite end, for energies of the order of $\mpl$ or larger, one presumably
needs a new quantum theory which includes gravity in a fundamental manner,
like String Theory~\cite{string} or Loop Quantum Gravity~\cite{loop}.
Or that realm is simply devoid of physical significance if the classicalization hypothesis
holds~\cite{dvali-a}.
\par
We hence expect that gravitational corrections to QFT amplitudes play an increasingly important
role for larger and larger energy scale $\mu>E_{\rm exp}$, and that it should be possible
to describe such effects in perturbative QFT directly (at least in the regime
$E_{\rm exp}\lesssim\mu\lesssim \mpl$).
We shall call this window the realm of ``semiclassical gravity'',
and that is the range where our proposal is more likely to shed some new light.
\subsection{Semiclassical gravity}
As we just mentioned, at intermediate energies $E_{\rm exp}\lesssim\mu\ll \mpl$, we expect that
a semiclassical picture holds in which the space-time can be reliably described as a classical
manifold with a metric tensor $g_{\alpha\beta}$ that responds to the presence of (quantum)
matter sources according to~\cite{birrell}
\be
R_{\alpha\beta}-\frac{1}{2}\,R\,g_{\alpha\beta}
=\frac{\lp}{\mpl}\,\expec{\hat T_{\alpha\beta}}
\ ,
\label{E}
\ee
where $R_{\alpha\beta}$ ($R$) is the Ricci tensor (scalar) and $\expec{\hat T_{\alpha\beta}}$
the expectation value of the matter stress tensor obtained from QFT on that background.
All the same, if one takes Eq.~\eqref{E} at face value, the way perturbative terms are computed in QFT
appears questionable, since loops of virtual particles are included in Feyman's diagrams
whose four-momentum $|k^2|=|k_\alpha\,k^\alpha|$ formally goes all the way to infinity
({\em i.e.}, to $\mpl$ and beyond), but are still described by the (free) propagators computed on a
fixed (and usually flat) background~\footnote{Note that we are qualifying divergences in
a relativistically covariant sense, by using the modulus of the four-momentum, like in Ref.~\cite{kempf}.}.
\par
\begin{figure}[t]
\centering
$k$
\\
$p_2$
\hspace{-0.5cm}
\raisebox{3.0cm}{$p_1$}
\epsfxsize=10cm
\epsfbox{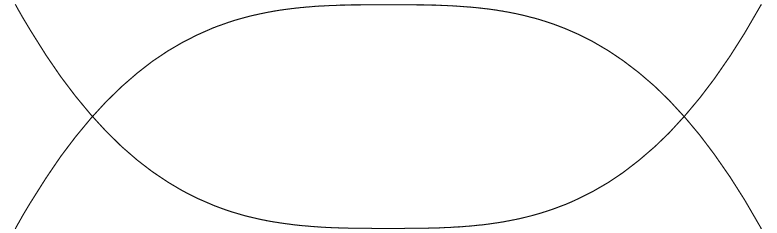}
$p_4$
\hspace{-0.5cm}
\raisebox{3.0cm}{$p_3$}
\\
$p_1+p_2-k$
\caption{One-loop correction to the four-point function in $\lambda\,\phi^4$.
\label{loop}}
\end{figure}
\begin{figure}[t]
\centering
\hspace{-3.5cm}$k$
\hspace{1.5cm}$k-q$
\hspace{0.5cm}$\cdots$
\\
$p_2$
\hspace{-0.5cm}
\raisebox{3.0cm}{$p_1$}
\epsfxsize=10cm
\epsfbox{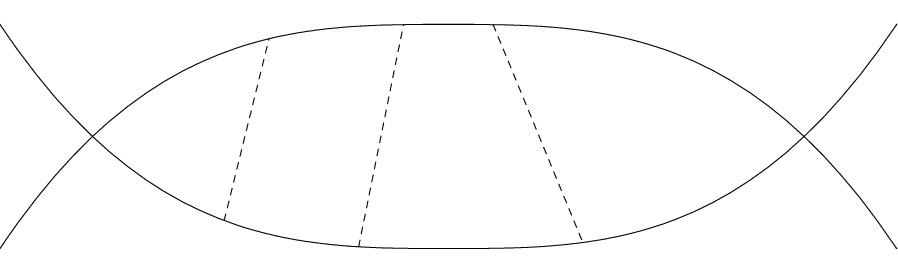}
$p_4$
\hspace{-0.5cm}
\raisebox{3.0cm}{$p_3$}
\\
\hspace{-4.5cm}
$p_1+p_2-k$
\hspace{0.2cm}$p_1+p_2+q-k$
\hspace{0.2cm}$\cdots$
\caption{One-loop correction to the four-point function in $\lambda\,\phi^4$
with graviton insertions (dashed lines).
\label{loopwg}}
\end{figure}

For example, let us consider the graph in Fig.~\ref{loop} for scalar particles with
self-interaction $\lambda\,\phi^4$, which is a pictorial representation of the
integral
\be
\Gamma^{(4)}(p)
\simeq
\int
\frac{k^3\,\ud k}{(2\,\pi)^4}\,
\tilde G_{\rm F}(k)\,\tilde G_{\rm F}(p-k)
\ ,
\label{G4}
\ee
where $\tilde G_{\rm F}$ is the momenutm-space Feynman propagator in four dimensions,
\be
\tilde G_{\rm F}(p)
=
\frac{1}{p^2+i\,\epsilon}
\ .
\label{GF}
\ee
Although the external momenta $|p_i|$ ($i=1,\ldots,4$) are taken within the range of
experiments (that is, $|p_i^2|=m_i^2\lesssim E_{\rm exp}^2$ in the laboratory frame),
the two virtual particles in the loop have unconstrained
momenta $k$ and $p_1+p_2-k$ respectively.
One might therefore wonder if it is at all consistent to describe those two particles
using the above flat-space propagator.
The common QFT approach to this problem would result in adding gravity in the form of graviton
exchanges (see Fig.~\ref{loopwg}) and estimate deviations from purely flat-space results.
This procedure will however not render finite diverging integrals, such as the one in Eq.~\eqref{G4},
unless one is able to resum an infinite number of perturbative terms.
\par
The interplay among propagators, UV~divergences and the causal structure of space-time 
can be better understood by noting that, in any approach in which the
space-time structure is given by a fixed background, the short distance behaviour of QFT
(in four dimensions) is described by the Hadamard form of the propagators~\cite{ford}
\be
G^{(2)}(x,x')
=
U(x;x')\,\delta(\sigma)
+V(x;x')\,\Theta(-\sigma)
\ ,
\ee
where $U$ and $V$ are regular functions and $2\sigma$ is the square
of the geodesic distance between $x$ and $x'$.
In Minkowski space-time,
\be
2\sigma=(x-x')^2
\ ,
\label{sigma}
\ee
and the expression of the propagator contains divergences for $\sigma\to 0$
({\em i.e.}, along the light cone and for $x\to x'$).
Calculations based on the use of propagators in QFT therefore (implicitly) rely on
the formalism of distribution theory and UV~divergences appear as a consequence
when one tries to compute (mathematically) ill-defined quantities such as the four-point
function in Eq.~\eqref{G4}.
One can devise mathematical workarounds for this problem, but
what matters here is that, if only the relation~\eqref{sigma} is modified
(like in QFT on a curved space-time), the divergences for $\sigma\to 0$ will remain,
because virtual high four-momentum modes remain essentially unaffected and lead to
the same ``overcountings'' as in flat space.
Nonetheless, a few partial results suggest that deeper modifications of the causal structure
might occur at the quantum level.
For example, it was shown that the divergence on the light-cone disappears
(with a smearing at large momenta of the form considered in Ref.~\cite{spallucci-a,spallucci-b}),
if graviton fluctuations are in a coherent state.
\par
It seems sensible to us to tackle this problem by pushing further the validity of  the
semiclassical Einstein equations.
We shall hence assume that virtual particles propagate
in a background compatible with Eq.~\eqref{E} at the scale $\mu\sim k=\sqrt{|k^2|}$
and their propagators be correspondingly adjusted~\cite{deser}.
As we mentioned in the Introduction, our underlying viewpoint here is that gravity is not
just another field (although with a very complicated dynamics), but the geometrical view
according to which gravity {\em is\/} the space-time and, in particular, the causal
structure obeyed by all (other) fields.
Let us remark again that this perspective is partly incorporated into the background
field method, whereby the metric is split into two parts,
\be
g_{\mu\nu}=\eta_{\mu\nu}+h_{\mu\nu}
\ .
\ee
The classical part $\eta_{\mu\nu}$ possesses the expected symmetries
of General Relativity and determines the causal structure for all (other) classical
and quantum fields, whereas $h_{\mu\nu}$ is just another quantum
field which acts on the matter fields via usual (although complicated) interaction
terms, hence in a non-geometrical way.
In comparison, one could actually view our approach as a step backward, since the
gravitational field is not explicitly quantised (there is no analogue of the above $h_{\mu\nu}$),
and it is in fact not even defined separately
({\em i.e.},~in the absence of matter~\footnote{This is reminiscent of the ``relational mechanics''
approach to gravity (see, {\em e.g.\/}, Ref.~\cite{anderson08} and References therein).}).
\subsection{Gravity in propagators}
\label{set}
Inspired by the previous considerations, we formulate the following basic
prescriptions for defining a ``gravitationally renormalised'' QFT:
\begin{description}
\item[A1)]
perturbative QFT defined by Feynman diagrams
is a viable approach to particle physics for four-momenta $\mu$
below a cut-off $\Lambda\gg E_{\rm exp}$;
\item[A2)]
in a (one-particle irreducible) Feynman diagram with $N$~internal lines, each
virtual particle is described by a Feynman propagator $G_{\{k_i\}}^{(\Lambda)}(x,y)$
corresponding to the space-time generated by the other $N-1$ virtual particles
in the same graph with momenta $k_i$ ($i=1,\ldots,N-1$) and constrained
according to~{\bf A1};
\item[A3)]
Standard Model results are recovered at low energy,
$\mu\lesssim E_{\rm exp}\ll\mpl$.
\end{description}
\par
Several comments on the above guidelines are in order.
First of all, we explicitly introduced a cut-off in {\bf A1}, having in mind
that we are not trying to build the final theory of everything,
but rather formulate a computational recipe.
It cannot indeed be excluded that our approach is just equivalent to more standard
perturbative QFT methods, and that our modified matter propagators just reproduce
the same effects obtained in the limit when infinitely many graviton exchanges
are included in graphs like the one in Fig.~\ref{loopwg}.
A second, essential, simplification was introduced in {\bf A2}, in that
each virtual particle is treated like a test particle in the space-time generated
by the other particles, its own gravitational backreaction thus being
neglected~\footnote{Let us note in passing that this somewhat parallels a
perturbative result of non-commutative QFT, according to which there is
no tree-level correction to the commutative case~\cite{szabo}.}.
Another consequence of {\bf A2} is that integration over momenta inside loops
can now be viewed as also purporting a (quantum mechanical) superposition of
(virtual) metrics, and there is hope that this can smear the usual divergences
of~\eqref{GF} out (as was shown in Ref.~\cite{ford} for particular gravitational states).
\par
It should not go unnoticed that we did not mention a Lagrangian (or action)
from which the modified propagators satisfying {\bf A2} could be obtained.
In this respect, our proposal follows the philosophy of Ref.~\cite{veltman},
which gives the Lagrangian a secondary role with respect to Feynman's
rules for computing perturbative amplitudes.
It is however true that the symmetries of a system are far more
transparent if a Lagrangian is available~\cite{weinberg97}, and
it would be interesting to find out whether an action principle can be
devised to streamline the derivation and show which symmetries
are preserved or broken.
The latter kind of analysis can also be performed perturbatively, although,
as is well known for the Slavnov-Taylor identities of (non-Abelian) Yang-Mills theory,
that task requires a lot more effort.
\par
A final observation is that the Standard Model of particle physics (without
gravity) is a rigid theory, and it is very likely that a generic modification of the
sort we are proposing here has hazardous effects in the range of presently available data,
thus compromising {\bf A3}.
One should therefore check very carefully that none of the assessed predictions
of the Standard Model is lost in our approach.
\par
At this point, we cannot proceed ignoring the technical fact that the $N$-body problem in
General Relativity is extremely complicated, to say the least, already for $N=2$.
We therefore make the following ``mean field'' assumption to deal with graphs
containing more than two virtual particles~\footnote{Although no such graph will
be considered here.}:
\begin{description}
\item[A4)]
one can approximate the propagator for each virtual particle
$G_{\{k_i\}}^{(\Lambda)}(x,y)\simeq G_{q}^{(\Lambda)}(x,y)$,
where $q\simeq \sqrt{|\sum k_i|^2}$ is the total momentum of the
remaining $N-1$ particles.
\end{description}
Since this is intended to be a (necessary) working assumption,
the above approximate equalities can be replaced with other expressions
of choice, the key point being that the problem is now reduced to study
the propagator for a test particle in a background generated by
an ``average'' source~\cite{shell}.
Again, it will be crucial that {\bf A3} remains valid in order to have a physically
sensible construction.
\section{Scalar~QFT}
\setcounter{equation}{0}
\label{scalar}
We shall now apply our prescriptions {\bf A1}--{\bf A4} to the simple case of a neutral
massless scalar field $\phi$ in four dimensions with $\lambda\,\phi^4$ self-interaction.
Although {\bf A3} (that is, the Standard Model physics) is lost from the outset, this model
is quite adequate for studying the UV~(or short distance) behaviour of QFT,
which is our main concern here.
\par
To begin with, we need a metric that describes the space-time around
virtual particles from which a propagator satisfying {\bf A2}
can be obtained.
Although {\bf A4} implies that we can reduce the problem to the relatively simple
case of one source, this is not yet enough to single out the metric to use,
and we shall need to make more working assumptions hereafter.
\subsection{Point-particle metrics}
Since we are considering neutral scalar particles,
the first option for a point-like source of mass $m$ that comes to mind is,
of course, the well-known Schwarzschild metric,
\be
\ud s^2
=
-\left(1-\frac{2\,\bar m}{\bar r}\right)\ud t^2+
\left(1-\frac{2\,\bar m}{\bar r}\right)^{-1}\ud \bar r^2+\bar r^2\,\ud\Omega^2
\ ,
\label{schwa}
\ee
where $\ud \Omega^2$ is the line element of the unit two-sphere and
$\bar r$ is the usual areal radius.
We also denoted with $\bar m=\lp\,m/\mpl$ the particle mass in geometric units,
which equals the ADM mass of the system, that is the asymptotic limit for $\bar r\to\infty$
of the (in this case, constant) mass function $\bar m=\bar m(\bar r)=\bar m\,\Theta(\bar r)$.
For $m>0$, the space-time with metric~\eqref{RN} represents a black hole.
However, the Compton length associated with a massive particle is
$\lp\,\mpl\,m^{-1}\gg \bar m$ for $m\ll \mpl$, and one might argue whether the horizon
really survives quantum mechanical corrections for $m$ comparable to the mass of
known elementary particles~\footnote{For example, space-time non-commutativity could
change the short distance metric into a de~Sitter-like background~\cite{nicolini}.}.
Moreover, and what is more important for our purpose of computing Feynman diagrams,
the causal structure of the Schwarzschild space-time would make the analytic expression
for the scalar field propagator overly complicated at short distances
(that is, right around $\bar r\sim \bar m$), although this is precisely the point of
view taken in Ref.~\cite{dvali-a}.
The Schwarzschild metric was also used in Ref.~\cite{adm},
in the so-called isotropic form
\be
\ud s^2
=
-\left(\frac{2\,r-\bar m}{2\,r+\bar m}\right)^2\ud t^2
+\left(1+\frac{\bar m}{2\,r}\right)^4\left(\ud r^2+r^2\,\ud\Omega^2\right)
\ ,
\label{RN}
\ee
which represents a wormhole, rather than a black hole, as the natural geometry
of ``point-like'' particles.
The point-like limit is effectively replaced by a configuration in which the
particle retains finite surface area and the ADM mass vanishes
(for more details, see Ref.~\cite{point}). 
This shows that gravity is indeed able to cure the point-like singularities of
classical field theory.
\par
We shall here take a step forward and assume the isotropic form can
be approximately extended from spatial coordinates to all of the four dimensions.
In order to see this, let us consider the coordinate transformation
\be
T= \left(\frac{2\,r-\bar m}{2\,r+\bar m}\right)\left(1+\frac{\bar m}{2\,r}\right)^{-2} t
\ ,
\ee
from which
\be
\left(\frac{2\,r-\bar m}{2\,r+\bar m}\right) \ud t
=
\left(1+\frac{\bar m}{2\,r}\right)^2 \ud T
-
\frac{\bar m\,T}{r^2}\left(2-\frac{\bar m}{2\,r}\right)
\left(\frac{2\,r+\bar m}{2\,r-\bar m}\right)
\ud r
\ .
\label{dT}
\ee
The second term diverges at the throat of the wormhole, $r=\bar m/2$
(corresponding to $\bar r=2\,\bar m$).
It was however shown in Ref.~\cite{point} that only a source with
$\mpl\lesssim m\lesssim 4\,\mpl$ extends less than the throat,
and such a configuration should be avoided for (virtual) particles with mass $m<\mpl$.
We further note that, for a virtual source of energy $m$, Heisenberg's uncertainty
principle implies a bound on the particle's lifetime $T$ such that
\be
\bar m\,T= \frac{\lp}{\mpl}\,m\,T
\sim
\frac{\lp}{\mpl}\,\lp\,\mpl
=
\lp^2
\ ,
\ee
or $T\sim \lp\,\mpl/m$, which is precisely the Compton length of the particle.
The metric elements in Eq.~\eqref{dT}, for virtual particles of mass $m\lesssim\mpl$,
can now be estimated assuming $T\sim \lp\,\mpl/m\sim r$ .
We respectively find
\be
\left(1+\frac{\bar m}{2\,r}\right)^2
\sim
\left(1+\frac{m^2}{2\,\mpl^2}\right)^2
\sim
1
\ ,
\ee
and
\be
\frac{\bar m\,T}{r^2}\left(2-\frac{\bar m}{2\,r}\right)
\left(\frac{2\,r+\bar m}{2\,r-\bar m}\right)
\sim
\frac{m^2}{\mpl^2}\left(2-\frac{m^2}{2\,\mpl^2}\right)
\left(\frac{2\,\mpl^2+m^2}{2\,\mpl^2-m^2}\right)
\sim
\frac{m^2}{\mpl^2}
\ll 1
\ .
\ee
which is negligible with respect to the previous term.
We can therefore approximate the metric with the significantly simpler conformally flat
form
\be
g_{\mu\nu}\simeq \Omega^2_m\,\eta_{\mu\nu}
\ ,
\label{Og}
\ee
where $\eta_{\mu\nu}$ is the Minkowski metric,
\be
\Omega_{m}(\vec x)=\left(1+\frac{\bar m}{2\,r}\right)^2
\ ,
\ee
and $r=|\vec x|$ is again the radial coordinate centered at the Dirac~$\delta(r)$ source
of bare mass $m$.
\subsection{Modified propagator}
Having chosen the metric in Eq.~\eqref{Og}, the next step is then to relate the
bare mass $\bar m$ with the momentum of the virtual particles regarded
as background sources.
According to {\bf A4}, we shall assume  
\be
m\simeq \sqrt{|q^2|}
\equiv q
\ ,
\ee
where $q$ is the either the total or the average momentum of $N-1$
virtual particles in a graph with $N$ such particles.
\par
We can now obtain the general form of the propagator starting from
the Klein-Gordon equation for the metric~\eqref{Og},
\be
\Box\,\phi=\Omega_q^{-3}\,\Box_{\rm M}\,(\Omega_q\,\phi)=0
\ ,
\ee
where $\Box_{\rm M}$ is the D'Alembertian in Minkowski space.
The modified Feynman propagator in coordinate space (with $x=(t,\vec x)$, etc)
is thus
\be
G_{q}^{(\Lambda)}(x,y)=
\Omega^{-1}_{q}(\vec x)\,G_{\rm F}(x-y)\,\Omega^{-1}_{q}(\vec y)
\ ,
\label{Gpx}
\ee
where $G_{\rm F}(x-y)$ is the standard Feynman propagator in Minkowski
space-time and the factors of $\Omega^{-1}_{q}$ are expected to suppress
the propagation of scalar modes at short distance, {\em i.e.},~for
$|\vec x|$, $|\vec y|\lesssim \lp\,q/\mpl$.
\par
\begin{figure}[t]
\centering
\epsfxsize=10cm
\epsfbox{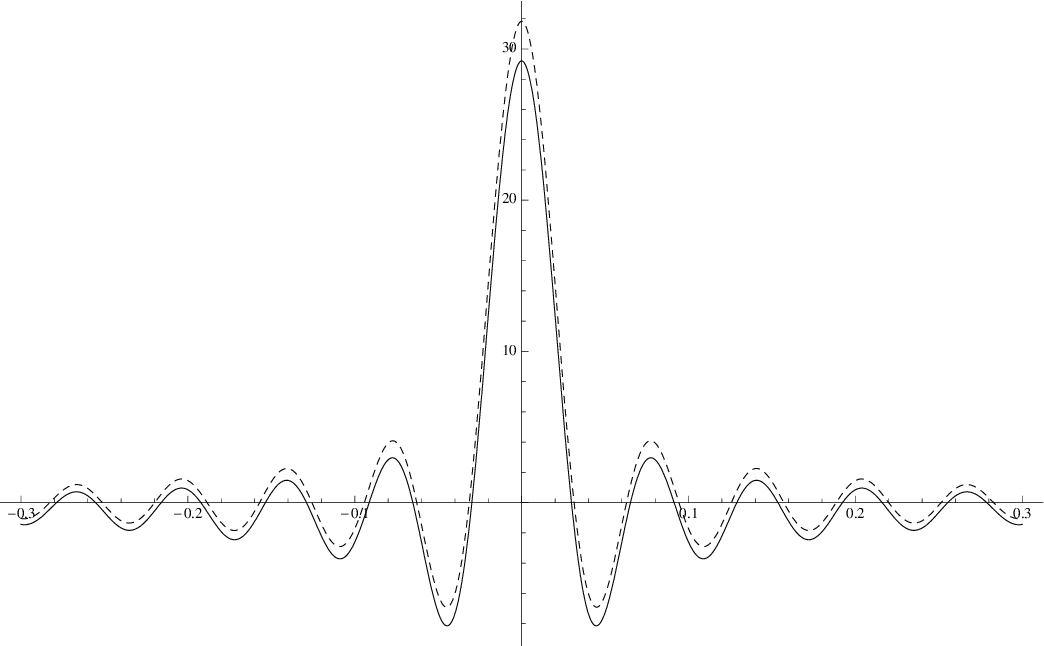}
\raisebox{0.5cm}{$\!\!\!\frac{w}{\mpl}$}
\caption{Distribution $\rho_{q}^{(L)}(w)$ for $q=\mpl$ (thick line)
and $q=0$ (dashed line) for $L=100\,\lp$.
\label{F1}}
\end{figure}

In order to see whether this improved behaviour is sufficient to cure
UV~divergences, we compute the propagator in momentum space
by taking the Fourier transform of~\eqref{Gpx}, with the cut-off $k<\Lambda$
according to {\bf A1},
\be
\tilde G_{q}^{(\Lambda)}(p;p')=
\int^\Lambda
\!\! 
\tilde\Omega_{q}(\vec p-\vec k)\,
\tilde G_{\rm F}(k)\,
\tilde\Omega_{q}(\vec k-\vec p')
\,\ud^3 k
\ ,
\ee
where $\tilde G_{\rm F}(k)$ is the standard
Feynman propagator in momentum space and
\be
\tilde \Omega_{q}(\vec k)
=
\frac{1}{(2\,\pi)^3}
\int
\frac{e^{-i\,\vec k\cdot\vec x}}{\Omega_{q}(\vec x)}\,\ud^3 x
\ .
\label{tO}
\ee
One can study this distribution as usual by integrating inside the box
$-\vec L<\vec x<\vec L$, and then taking $\vec L\to\infty$.
By rotating the reference frame so that $\vec k=(k_x,0,0)$, we find
\be
\tilde \Omega_{q}(\vec k)
=
\delta(k_y)\,\delta(k_z)
\lim_{L_x\to\infty}
\rho^{(L_x)}_{q}(k_x)
\ ,
\ee
where $\delta(w)$ is the Dirac $\delta$-function and
\be
\rho^{(L)}_{q}(w)=
\frac{1}{2\,\pi}
\int_{-L}^{+L}\!\!
\frac{x^2\,e^{-i\,w\,x}\,\ud x}
{(|x|+\lp\,q/\mpl)^2}
\ .
\label{tOL}
\ee
Its explicit expression is rather cumbersome,
\be
\rho^{(L)}_{q}(w)=
2\,\bar q
+e^{-i\,w\,\bar q}\,\bar q\,\left(2-i\,k\,\bar q\right)\left[
{\rm Ei}(i\,k\,\bar q)-{\rm Ei}(i\,k\,(L+\bar q))\right]
+i\,e^{-i\,k\,L}\frac{L+\bar q+i\,k\,\bar q^2}{k\left(L+\bar q\right)}
+{\rm c.c.}
\ ,
\ee
where c.c.~stands for complex conjugate,
\be
{\rm Ei}(z)=\int_{-\infty}^z t^{-1}\,e^t\,\ud t
\ ,
\ee
and $\bar q=(\lp/\mpl)\,q$.
Upon closer inspection, one realises that $\rho^{(L)}_{q}(w)$
is real and even in $w$, and actually
resembles the usual approximation of $\delta(w)$ (see Fig.~\ref{F1}),
with $\rho_{q}^{(L)}(0)\simeq L$ and the normalisation
\be
\lim_{L\to\infty}\int_{-\Lambda}^{+\Lambda}
\rho_{q}^{(L)}(w)
\,\ud w
=
\rho_\Lambda(q)
\ ,
\ee
where $\Lambda$ is again the cut-off introduced in {\bf A1\/}.
Finally, we obtain
\be
\tilde \Omega_{q}(\vec k)
=
\rho_\Lambda(q)\,\delta(\vec k)
\ ,
\ee
and the relevant propagator is therefore given by
\be
\tilde G_{q}^{(\Lambda)}(p)
=
\rho^2_\Lambda(q)\,
\tilde G_{\rm F}(p)
\ ,
\label{Gp}
\ee
in which the weight $\rho_\Lambda^2$ describes, in momentum space, the previously
mentioned suppression at short distance.
Remarkably, an explicit dependence on the UV cut-off $\Lambda$ and $\mpl$ emerged
as manifestation of non-trivial IR/UV mixing at all scales $p\sim\mu>0$.
\par
It is interesting to note that the limit
\be
\lim_{\Lambda\to\infty}\rho_\Lambda^2(q)=
\left\{
\begin{array}{ll}
1
& 
{\rm for}\ q=0
\\
\\
0
&
{\rm for}\ 
q>0
\ ,
\end{array}
\right.
\label{rhosing}
\ee
which is therefore not uniform and must be taken carefully at the end of the
computation only.
Further, since one has $\rho_\Lambda(0)=1$ for all values of $\Lambda$,
the standard propagator~\eqref{GF} (with no dependence on $\Lambda$ and $\mpl$)
is recovered for $q/\mpl\to 0$ followed by $\Lambda\to\infty$.
One can therefore approximate
\be
\tilde G_{(q\ll\mpl)}^{(\Lambda\gg p)}(p)
\simeq
\tilde G_{\rm F}(p)
\ ,
\ee
if need be.
\par
\begin{figure}[t]
\centering
\raisebox{5.0cm}{$\rho_{\Lambda}^2(q)$}
\epsfxsize=10cm
\epsfbox{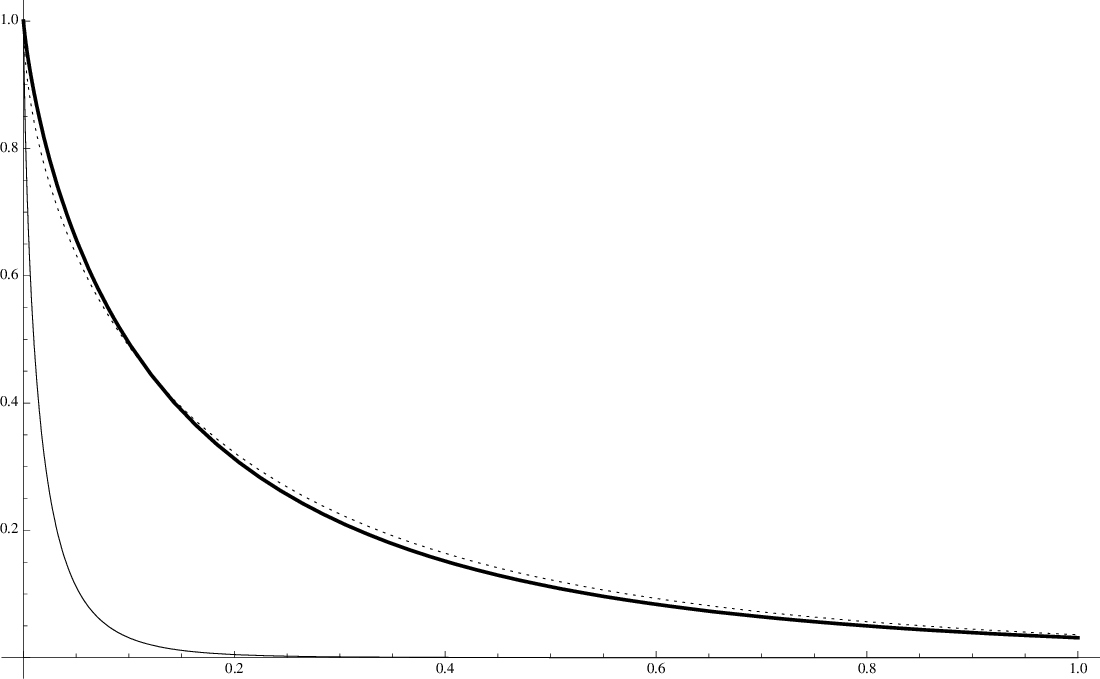}
${q}/\mpl$
\caption{Weight $[\rho^{(L)}_{\Lambda}(q)]^2$ with $L=10^6\,\lp$
for $\Lambda=10\,\mpl$ (thin solid line), $\Lambda=\mpl$ (thick solid line)
and its approximation~\eqref{rhop} (dotted line) with $\alpha=0.55$.
\label{F2}}
\end{figure}
In order to study the UV~behaviour of transition amplitudes, we shall need
analytically manageable approximations of the propagator~\eqref{Gp} for
$q\lesssim \Lambda$.
For $(\mpl^2/\Lambda)\lesssim q<\Lambda$, we numerically find the bound
\be
\rho_{\Lambda\gg\mpl}^2(q)
<
\left(\frac{\mpl^2}{\Lambda\,q}\right)^\beta
\ ,
\label{limit}
\ee
with $\beta\simeq 5.8$,
which can be used to estimate quantities in the limit $\Lambda\to\infty$.
It is also tempting to relate $\Lambda$ to $\mpl$ explicitly~\cite{salam}.
In this case, we numerically checked that the weight can be approximated by 
\be
\rho_{\Lambda\simeq\mpl}^2(q\lesssim \mpl)
\simeq
1-\tanh\left[2\left(\frac{\Lambda\,q}{\mpl^2}\right)^{\alpha}\right]
\ ,
\label{rhop}
\ee
with $\alpha\simeq 0.55$ (see Fig.~\ref{F2}).
Note that neither~\eqref{limit} nor~\eqref{rhop} is accurate for $q\to 0$ and
we are thus not providing useful approximations to study the IR behaviour.
\subsection{Scattering amplitudes}
From Eq.~\eqref{limit}, one already suspects that the propagator~\eqref{Gp} yields finite
amplitudes for all the irreducible diagrams involving at least two virtual particles.
Let us see this in detail for the one-loop correction to the vertex $\lambda\,\phi^4$ of
Fig.~\ref{loop}.
\par
The standard asymptotic behaviour for the total momentum of
the incoming scalars $p\ll\Lambda$ is
\be
\Gamma^{(4)}(p)
\simeq
\int^\Lambda
\!\!
\frac{k^3\,\ud k}{(2\,\pi)^4}\,
\tilde G_{\rm F}(k)\,\tilde G_{\rm F}(p-k)
\simeq
C\,\ln\left(\frac{\Lambda}{p}\right)
\ ,
\label{G4sm}
\ee
with $C$ a constant of order one.
This amplitude contains the formal (that is, bare) coupling constant $\lambda$,
which can be replaced with the physical coupling constant given by the transition amplitude
measured at the scale $\mu$,
\be
\lambda_\mu\simeq\lambda+\lambda^2\,\Gamma^{(4)}(\mu)
\ .
\ee
Upon solving for $\lambda$ in terms of $\lambda_\mu$,
the scattering amplitude for $p\sim\mu$ becomes a function of
the two physically meaningful quantities $\lambda_\mu$ and $\mu$,
\be
\mathcal{M}
\simeq
\lambda_\mu-\lambda_\mu^2\,C\,
\ln\left(\frac{p}{\mu}\right)
\ .
\label{M2}
\ee
This expression is independent of $\Lambda$, so that
the low energy physics ($p\sim\mu\ll\Lambda$) depends
on the (otherwise unknown) high energy theory ($k\gtrsim\Lambda$)
only through the ``renormalized'' value of $\lambda_\mu$.
The UV~cut-off may then be removed safely (formally, taking $\Lambda\to\infty$
at the end of the computation does not affect the result).
\par
The ``gravitationally renormalized'' amplitude is obtained by replacing
each particle's propagator in Eq.~\eqref{G4sm} with the expression~\eqref{Gp}
and $q$ equal to the momentum of the other virtual particle,
\be
\Gamma^{(4)}_{\rm GR}(p)
\simeq
\int^\Lambda
\frac{k^3\,\ud k}{(2\,\pi)^4}\,
\tilde G_{(p-k)}^{(\Lambda)}(k)\,\tilde G_{(k)}^{(\Lambda)}(p-k)
\ .
\label{G4gr}
\ee
The result now depends on $\Lambda$ and we shall consider two cases by
making use of the approximate expressions for $\rho^2_\Lambda$ found
previously.
\subsubsection{Infinite cut-off}
It is easy to see that the asymptotic behaviour (for $p\ll\Lambda\to\infty$)
is now given by
\be
\Gamma^{(4)}_{\rm GR}(p)
\lesssim
\frac{\mpl^{4\beta}}{\Lambda^{2\beta}}
\int^\Lambda
\!\!
\frac{\ud k}{k^{1+2\beta}}
\sim
\left(\frac{\mpl}{\Lambda}\right)^{4\beta}
\ ,
\ee
and the integral~\eqref{G4gr} therefore remains finite for $\Lambda\to\infty$.
The same occurs in all higher order cases and the only
irreducible graph left (potentially~\footnote{In the Standard Model, this would
only occur for the gluon self-mass~\cite{gogokhia}.})
diverging is the tadpole, since it only contains one virtual particle
propagated by $\tilde G_{(q=0)}^{(\Lambda)}(k)=\tilde G_{\rm F}(k)$.
\subsubsection{Planck scale cut-off}
If we instead identify $\Lambda\simeq\mpl$~\cite{salam}, we obtain
\be
\Gamma^{(4)}_{\rm GR}(p)
\simeq
C\,\ln\left(\frac{\mpl}{p}\right)
\ ,
\label{fprop}
\ee
which shows an explicit dependence on $\mpl$ as anticipated.
Incidentally, this is exactly the same asymptotic behaviour one finds from the
standard expression~\eqref{G4sm} by simply setting $\Lambda=\mpl$,
which shows that deviations from purely Standard Model UV~results
are only obtained by pushing the cut-off above the Planck scale.
\par
Let us repeat that the displayed results only pertain to the high energy regime and,
to complete the analysis, one should also consider the IR~behaviour more carefully.
\section{Final remarks}
\setcounter{equation}{0}
Inspired by the observation that a semiclassical description of gravity
should be possible in processes that involve energies below the Planck scale,
we formulated general properties that a modified QFT should enjoy in order to include
gravitational contributions.
Such properties were listed in the form of prescriptions that formalise
our approach to include gravity within the Standard Model of particle physics
in a non-perturbative way.
As such, they are of course debatable and subject to possible refinements.
\par
In order to have a first look at what predictions such guidelines imply, 
we then derived the Feynman propagator~\eqref{Gp} for a neutral
scalar field.
However, to carry on the computation analytically required several more working
assumptions, starting from the choice of the background metric~\eqref{Og}.
The resulting propagator explicitly depends on the energy (length) scale $\mpl$ ($\lp$)
and cut-off $\Lambda$, which entails a IR/UV mixing, with the high energy scale
$\Lambda$ (possibly proportional to $\mpl$) that appears explicitly in the low energy
scattering amplitudes.
From the phenomenological point of view, our approach can therefore be regarded
as an attempt to predict the effects of the existence of a fundamental length
in QFT~\footnote{The physical value of the cut-off $\Lambda$ could then be estimated by
(high precision) measurements such as the electron or muon $g-2$.}.
Results such as~\eqref{rhop} and~\eqref{fprop} are consequently
illustrative of the magnitude of the UV~gravitational corrections one expects
in four space-time dimensions, where $\mpl\gg E_{\rm exp}$ and we know
{\em a priori\/} that it all must boil down to very small figures~\footnote{The situation
might be remarkably different in models with extra-spatial dimensions and
$\mpl\simeq 1\,$TeV~\cite{extra-a,extra-b,extra-c,extra-d,extra-e,lbh}.}.
\par
As for the long-standing problem of the UV~behaviour of QFT, we
need to push our semiclassical scheme by letting $\Lambda\gg\mpl$
(like in Ref.~\cite{robinson}) in order to tackle it.
Our conclusion using~\eqref{Gp} is that the dependence on the UV
cut-off is much improved over that of the standard QFT propagators
and finite results without the need of removing divergences are expected in all
cases but the few involving just one virtual particle
(like the tadpole diagram for a scalar field).
We cannot, however, exclude that the asymptotic behaviour might
change again by considering more refined descriptions.
For instance, one should likely relax sphericity and conformal
flatness~\cite{reuter08} of the metric~\eqref{Og},
since these hardly suit systems of particles with large relative
momenta.
And, of course, more realistic QFT should be analysed before the final word
can be spoken on that old idea of Pauli.
\end{document}